\documentclass[aps,floats,floatfix,showpacs,twoside,preprint,superscriptaddress]{revtex4}

\usepackage{latexsym}
\usepackage{amsfonts}
\usepackage{graphicx}
\usepackage{amsbsy}
\usepackage{float}

\begin{document}

\title{Quantifying and Modeling Long-Range Cross-Correlations  in 
 Multiple Time Series with Applications to World Stock Indices} 

\author{Duan Wang}
\affiliation{Center for Polymer Studies and Department of Physics, Boston
University, Boston, MA 02215, USA}

\author{Boris~Podobnik}
\affiliation{Center for Polymer Studies and Department of Physics, Boston
University, Boston, MA 02215, USA}
\affiliation{Faculty of Civil Engineering, University of Rijeka, 51000
Rijeka, Croatia}
\affiliation{Zagreb School of Economics, 10000 Zagreb, Croatia}

\author{Davor Horvati\'c}
\affiliation{Physics Department, Faculty of Science, University of Zagreb,
10000 Zagreb, Croatia}

\author{H.~Eugene~Stanley}
\affiliation{Center for Polymer Studies and Department of Physics, Boston
University, Boston, MA 02215, USA}

\begin{abstract} 

We propose a modified time lag random matrix theory in order to study time lag cross-correlations in multiple time series. We apply the method to 48 world indices, one for
each of 48 different countries.  We find long-range power-law
cross-correlations in the absolute values of returns that quantify risk,
and find that they decay much more slowly than cross-correlations
between the returns.  The magnitude of the cross-correlations constitute
``bad news'' for international investment managers who may believe that
risk is reduced by diversifying across countries. We find that when a
market shock is transmitted around the world, the risk decays very
slowly.  We explain these time lag cross-correlations by introducing a
global factor model (GFM) in which all index returns fluctuate in
response to a single global factor. For each pair of individual time
series of returns, the cross-correlations between returns (or
magnitudes) can be modeled with the auto-correlations of the global
factor returns (or magnitudes). We estimate the global factor using
principal component analysis, which minimizes the variance of the
residuals after removing the global trend. Using random matrix theory, a
significant fraction of the world index cross-correlations can be
explained by the global factor, which supports the utility of the GFM.
We demonstrate applications of the GFM in forecasting risks at the world
level, and in finding uncorrelated individual indices. We find 10
indices are practically uncorrelated with the global factor and with the
remainder of the world indices, which is relevant information for world
managers in reducing their portfolio risk. Finally, we argue that this
general method can be applied to a wide range of phenomena in which time
series are measured, ranging from seismology and physiology to
atmospheric geophysics.
 
\end{abstract}


\maketitle

\section{Introduction}
\label{intro}

When complex systems join to form even more complex systems, the
interaction of the constituent subsystems is highly random
\cite{havlin1,chris,havlin2,fn}. The complex stochastic interactions
among these subsystems are commonly quantified by calculating the
cross-correlations. This method has been applied in systems ranging from
nanodevices \cite{Samuelsson,Cottet,Nader}, atmospheric geophysics
\cite{Yama}, and seismology \cite{Seismology,Corral,Lipp}, to finance
\cite{Solnik96,Erb,LeBaron,prl,takayasubook,Mant99,Kertesz,Mantegna06,Takayasu06,DCCA,Carbone,BP07}.
Here we propose a method of estimating the most significant component in
explaining long-range cross-correlations.
           
Studying cross-correlations in these diverse physical systems provides
insight into the dynamics of natural systems and enables us to base our
prediction of future outcomes on current information. In finance, we
base our risk estimate on cross-correlation matrices derived from asset
and investment portfolios \cite{Mark,prl}. In seismology,
cross-correlation levels are used to predict earthquake probability and
intensity \cite{Seismology}. In nanodevices used in quantum information
processing, electronic entanglement necessitates the computation of
noise cross-correlations in order to determine whether the sign of the
signal will be reversed when compared to standard devices
\cite{Samuelsson}.  Reference~\cite{bob10} reports that
cross-correlations for $\Delta t=0$ calculated between pairs of EEG time
series are inversely related to dissociative symptoms (psychometric
measures) in 58 patients with paranoid schizophrenia.  In genomics data,
Ref.~\cite{Podobnik10} reports spatial cross-correlations corresponding
to a chromosomal distance of $\approx 10$ million base pairs.  In
physiology, Ref.~\cite{Podobnik10} reports a statistically significant
difference between alcoholic and control subjects.
 
Many methods have been used to investigate cross-correlations (i)
between pairs of simultaneously recorded time series \cite{DCCA,Carbone}
or (ii) among a large number of simultaneously-recorded time series
\cite{prl,Jolliffe86,Guhr03}. Reference~\cite{Guhr03} uses a power
mapping of the elements in the correlation matrix that suppresses noise.
Reference~\cite{DCCA} proposes detrended cross-correlation analysis
(DCCA), which is an extension of detrended fluctuation analysis (DFA) 
 \cite{dfa} 
and is based on detrended covariance.  Reference~\cite{Carbone} proposes
a method for estimating the cross-correlation function $C_{xy}$ of
long-range correlated series $x_t$ and $y_t$.  For fractional Brownian
motions with Hurst exponents $H_1$ and $H_2$, the asymptotic expression
for $C_{xy}$ scales as a power of $n$ with exponents $H_1$ and $H_2$.
   
Univariate (single) financial time series modeling has long been a
popular technique in science.  To model the auto-correlation of
univariate time series, traditional time series models such as
autoregressive moving average (ARMA) models have been proposed
\cite{Box70}. The ARMA model assumes variances are constant with
time. However, empirical studies accomplished on financial time series
commonly show that variances change with time.  To model time-varying
variance, the autoregressive conditional heteroskedasticity (ARCH) model
was proposed \cite{Engle82}. Since then, many extensions of ARCH has
been proposed, including the generalized autoregressive conditional
heteroskedasticity (GARCH) model \cite{Boll86} and the
fractionally-integrated autoregressive conditional heteroskedasticity
(FIARCH) model \cite{Granger96}. In these models, long-range
auto-correlations in magnitudes exist, so a large price change at one
observation is expected to be followed by a large price change at the
next observation.  Long-range auto-correlations in magnitude of signals
have been reported in finance \cite{Granger96}, physiology
\cite{Ashk,Kant}, river flow data \cite{Livina}, and weather data
\cite{PREBP05}.

Besides univariate time series models, modeling correlations in multiple
time series has been an important objective because of its practical
importance in finance, especially in portfolio selection and risk
management \cite{Sharpe64,Sharpe70}.  In order to capture potential
cross-correlations among different time series, models for coupled
heteroskedastic time series have been introduced
\cite{Boll88,Boll90,Engel95}.  However, in practice, when those models
are employed, the number of parameters to be estimated can be quite
large.
 
A number of researchers have applied multiple time series analysis to
world indices, mainly in order to analyze zero time-lag
cross-correlations.  Reference~\cite{Solnik96} reported that for
international stock return of nine highly-developed economies, the
cross-correlations between each pair of stock returns fluctuate strongly
with time, and increase in periods of high market volatility. By
volatility we mean time-dependent standard deviation of return.  The
finding that there is a link between zero time lag cross-correlations
and market volatility is ``bad news'' for global money managers who
typically reduce their risk by diversifying stocks throughout the world.
In order to determine whether financial crises are short-lived or
long-lived, Ref.~\cite{Canarella} recently reported that, for six Latin
American markets, the effects of a financial crisis are
short-range. Between two and four months after each crisis, each Latin
American market returns to a low-volatility regime.
 
In order to determine whether financial crisis are short-term or
long-term at the world level, we study 48 world indices, one for each of
48 different countries. We analyze cross-correlations among returns and
magnitudes, for zero and non-zero time lags.  We find that
cross-correlations between magnitudes last substantially longer than
between the returns, similar to the properties of auto-correlations in
stock market returns \cite{Ding93}. We propose a general method in order
to extract the most important factors controlling cross-correlations in
time series.  Based on random matrix theory \cite{prl} and principal
component analysis \cite{Jolliffe86} we propose how to estimate the
global factor and the most significant principal components in
explaining the cross-correlations. This new method has a potential to be
broadly applied in diverse phenomena where time series are measured,
ranging from seismology to atmospheric geophysics.

This paper is organized as follows.  In Section II we introduce the data
analyzed, and the definition of return and magnitude of return.  In
Section III we introduce a new modified time lag random matrix theory
(TLRMT) to show the time-lag cross-correlations between the returns and
magnitudes of world indices. Empirical results show that the
cross-correlations between magnitudes decays slower than that between
returns. In Section IV we introduce a single global factor model to
explain the short- or long-range correlations among returns or
magnitudes. The model relates the time-lag cross-correlations among
individual indices with the auto-correlation function of the global
factor. In Section V we estimate the global factor by minimizing the
variance of residuals using principal component analysis (PCA), and we
show that the global factor does in fact account for a large percentage
of the total variance using RMT. In Section VI we show the applications
of the global factor model, including risk forecasting of world economy,
and finding countries who have most the independent economies.

\section{Data Analyzed}
\label{data}

In order to estimate the level of relationship between individual stock
markets---either long-range or short-range cross-correlations exist at
the world level---we analyze $N=48$ world-wide financial indices,
$S_{i,t}$, where $i=1,2,\ldots,48$ denotes the financial index and $t$
denotes the time.  We analyze one index for each of 48 different
countries: 25 European indices \cite{europa}, 15 Asian indices
(including Australia and New Zealand) \cite{asia}, 2 American indices
\cite{amer}, and 4 African indices \cite{africa}. In studying 48
economies that include both developed and developing markets we
significantly extend previous studies in which only developed economies
were included---e.g., the seven economies analyzed in
Refs.~\cite{Longin95,Erb},
 and the 17 countries studied in Ref.~\cite{Ramchand}.
We use daily stock-index data taken from {\sl Bloomberg}, as opposed to
weekly \cite{Ramchand}
or monthly data  \cite{Solnik96}. 
The data cover the period 4 Jan 1999 through 10 July 2009, 2745 trading
days.  For each index $S_{i,t}$, we define the relative index change
(return) as
\begin{equation}
R_{i,t} \equiv\log S_{i,t} - \log S_{i,t-1},
\label{grang}
\end{equation}
where $t$ denotes the time, in the unit of one day.  By
magnitude of return we denote the absolute value of return after
removing the mean
\begin{equation}
|r_{i,t}|\equiv |R_{i,t}-\langle R_{i,t}\rangle|.
\end{equation}


\section{Modified Time-lag Random Matrix Theory}
\label{RMT}

\subsection{Basic Ideas of Time-lag Random Matrix Theory}
 
In order to quantify the cross-correlations, random matrix theory
({\small RMT}) (see Refs.~\cite{Mehta91} \cite{guhr} and references
therein) was proposed in order to analyze collective phenomena in
nuclear physics.  Refs.~\cite{prl} extended RMT to cross-correlation
matrices in order to find cross-correlations in collective behavior of
financial time series.  The largest eigenvalue $\lambda_+$ and smallest
eigenvalue $\lambda_-$ of the Wishart matrix \textsf{W} (a correlation
matrix of uncorrelated time series with finite length) are
\begin{equation}
\lambda_{\pm}=1+\frac{1}{Q} \pm 2 \sqrt{\frac{1}{Q}},
\label{rmt}
\end{equation}
where $Q\equiv T/N (>1)$, and $N$ is the matrix dimension and $T$ the
length of each time series.  The larger the discrepancy between (a) the
correlation matrix \textsf{C} between empirical time series and (b) the
Wishart matrix \textsf{W} obtained between uncorrelated time series, the
stronger are the cross-correlations in empirical data \cite{prl}.
Many {\small RMT} studies reported equal-time (zero $\Delta t$)
cross-correlations between different empirical time series
\cite{prl,laloux99b,Utsugi,pan,shen}.
  
Recently time-lag generalizations of {\small RMT} have been proposed
\cite{Pott05,new2,Bou07}.
In one of the generalizations of {\small RMT}, based on the eigenvalue
spectrum called time-lag {\small RMT} ({\small TLRMT}), Ref.~\cite{Podobnik10} 
found long-range cross-correlations in time series of price fluctuations
in absolute values of 1340 members of the New York Stock Exchange
Composite, in both healthy and pathological physiological time series,
and in the mouse genome.

We compute for varying time lags $\Delta t$ the largest singular values
$\lambda_L(\Delta t)$ of the cross-correlation matrix of N-variable time
series $X_{i,t}$
\begin{equation}
 C_{ij}(\Delta t) \equiv \frac{\langle X_{i,t}X_{j,t+\Delta t}\rangle
 -\langle X_{i,t}\rangle \langle X_{j,t+\Delta t}\rangle} {\sigma_i
 \sigma_j}.
 \label{C} 
\end{equation} 
We also compute $\tilde \lambda_L(\Delta t)$ of a similar matrix $\tilde
C(\Delta t)$, where $X_{i,t}$ are replaced by the magnitudes
$|X_{i,t}|$.  The squares of the non-zero singular values of \textsf{C} are
equal to the non-zero eigenvalues of \textsf{CC}$^+$ or \textsf{C}$^+$\textsf{C}, where by
\textsf{C}$^+$ we denote the transpose of \textsf{C}. In a singular value
decomposition ({\small SVD}) \cite{Gupta,Bou07,Podobnik10} 
 $\textsf{C} = \textsf{UDV}^+$ the diagonal elements of \textsf{D} are
equal to singular values of \textsf{C}, where the \textsf{U} and
\textsf{V} correspond to the left and right singular vectors of the
corresponding singular values.  We apply SVD to the correlation matrix
for each time lag and calculate the singular values, and the dependence
of the largest singular value $\lambda_L(\Delta t)$ on $\Delta t$ serves
to estimate the functional dependence of the collective behavior of
$C_{ij}$ on $\Delta t$ \cite{Podobnik10}.

\subsection{Modifications of Cross-Correlation Matrices}

We make two modifications of correlation matrices in order to better
describe correlations for both zero and non-zero time lags.

\begin{itemize}

\item[{(i)}] The first modification is a correction for correlation
  between indices that are not frequently traded. Since different
  countries have different holidays, all indices contain a large number
  of zeros in their returns. These zeros lead us to underestimate the
  magnitude of the correlations. To correct for this problem, we define
  a modified cross-correlation between those time series with extraneous
  zeros,
\begin{equation}
 C_{ij}'(\Delta t) \equiv \frac{1}{T'} \frac{\sum_{i=1}^T {X_{i,t}X_{j,t+\Delta t}}
 -\sum_{i=1}^T X_{i,t}  \sum_{i=1}^T X_{j,t+\Delta t}} {\sigma_i
 \sigma_j}.
 \label{C_adj} 
\end{equation}
Here $T'$ is the time period during which both $X_{i,t}$ and
$X_{j,t+\Delta t}$ are non-zero. With this definition, the time periods
during which $X_{i,t}$ or $X_{j,t+\Delta t}$ exhibit zero values have
been removed from the calculation of cross-correlations. The
relationship between $C_{ij}'(\Delta t)$ and $C_{ij}(\Delta t)$ is
\begin{equation}
C_{ij}'(\Delta t)=\frac{T}{T'}C_{ij}(\Delta t).
\end{equation}

\item[{(ii)}] The second modification corrects for
  auto-correlations. The main diagonal elements in the correlation
  matrix are ones for zero-lag correlation matrices and
  auto-correlations for non-zero lag correlation matrices. Thus,
  time-lag correlation matrices allow us to study both auto-correlations
  and time-lag cross-correlations. If we study the decay of the largest
  singular value, we see a long-range decay pattern if there are
  long-range auto-correlations for some indices but no cross-correlation
  between indices. To remove the influence of auto-correlations and
  isolate time-lag cross-correlations, we replace the main diagonals by
  unity,
\begin{equation}
C_{ij}''(\Delta t)=\cases{1        & when ~~ $i=j$ \cr
                          C_{ij}'(\Delta t) & when ~~ $i\ne j$}.
\end{equation}
With this definition the influence of auto-correlations is
  removed, and the trace is kept the same as the zero time-lag
  correlation matrix.

\end{itemize}

\subsection{Empirical Results} 

In Fig.~1(a) we show the distribution of cross-correlations between zero
and non-zero lags. For $\Delta t = 0$ the empirical pdf $P(C_{ij})$ of
the cross-correlation coefficients $C_{ij}$ substantially deviates from
the corresponding pdf $P(W_{ij})$ of a Wishart matrix, implying the
existence of equal-time cross-correlations.

In order to determine whether short-range or long-range
cross-correlations accurately characterize world financial markets, we
next analyze cross-correlations for $(\Delta t \ne 0)$.  We find that
with increasing $\Delta t$ the form of $P(C_{ij})$ quickly approaches
the pdf $P(W_{ij})$, which is normally distributed with zero mean and
standard deviation $1/\sqrt{N}$ \cite{Shumway}.

In Fig.~1(b) we also show the distribution of cross-correlations between
{\it magnitudes}.  In financial data, returns $R_{i,t}$ are generally
uncorrelated or short-range auto-correlated, whereas the magnitudes are
generally long-range auto-correlated \cite{Granger96,Liu}.
We thus examine the cross-correlations $\tilde C_{ij}(\Delta t)$ between
$|r_{i,t}|$ for different $\Delta t$. In Fig.~1(b) we find that with
increasing $\Delta t$, $P(\tilde C_{ij})$ approaches the pdf of random
matrix $P(W_{ij})$ more slowly than $P(C_{ij}) $, implying that
cross-correlations between index magnitudes persist longer than
cross-correlations between index returns.
      
In order to demonstrate the decay of cross-correlations with time lags,
we apply modified TLRMT. Fig.~2 shows that with increasing $\Delta t$
the largest singular value calculated for $\tilde {\textsf{C}}$ decays
more slowly than the largest singular value calculated for
\textsf{C}. This result implies that among world indices, the
cross-correlations between magnitudes last longer than
cross-correlations between returns. In Fig.~2 we find that $\lambda_L$
vs. $\Delta t$ decays as a power law function with the scaling exponent
equal to 0.25.  The faster decay of $\lambda_L$ vs. $\Delta t$ for
\textsf{C} implies very weak (or zero) cross-correlations among
world-index returns for larger $\Delta t$, which agrees with the
empirical finding that world indices are often uncorrelated in returns.
Our findings of long-range cross-correlations in magnitudes among the
world indices is, besides a finding in Ref.~\cite{Solnik96},
another piece of ``bad
news'' for international investment managers. World market risk decays
very slowly.  Once the volatility (risk) is transmitted across the
world, the risk lasts a long time. 

\section{Global Factor Model}
\label{SGFM}

The arbitrage pricing theory states that asset returns follow a linear
combination of various factors \cite{Ross}. We find that the factor
structure can also model time lag pairwise cross-correlations between
the returns and between magnitudes.  To simplify the structure, we model
the time lag cross-correlations with the assumption that each individual
index fluctuates in response to one common process, the ``global
factor'' $M_t$,
\begin{eqnarray}
R_{i,t}=\mu_i+b_i M_t+\epsilon_{i,t}. 
 \label{duan1}
\end{eqnarray}
Here in the global factor model (GFM), $\mu_i$ is the average return for
index $i$, $M_t$ is the global factor, and $\epsilon_{i,t}$ is the
linear regression residual, which is independent of $M_t$, with mean
zero and standard deviation $\sigma_i$. Here $b_i$ indicates the
covariance between $R_{i,t}$ and $M_t$, ${\rm Cov}(R_{i,t},M_t)=b_i {\rm
  Var}(M_t)$.  This single factor model is similar to the Sharpe market
model \cite{Sharpe63}, but instead of using a known financial index as
the global factor $M_t$, we use factor analysis to find $M_t$, which we
introduce in the next section. We also choose $M_t$ as a zero-mean
process, so the expected return $E(R_{i,t})=\mu_i$, and the global
factor $M_t$ is only related with market risk. We define a zero-mean
process
 $r_{i,t}$ as
\begin{equation}
  r_{i,t}\equiv R_{i,t}-E(R_{i,t})=b_i M_t+\epsilon_{i,t}.
\end{equation}
A second assumption is that the global factor can account for most of
the correlations. Therefore we can assume that there are no correlations
between the residuals of each index, ${\rm
  Cov}(\epsilon_{i,t},\epsilon_{j,t})=0$.  Then the covariance between
$R_{i,t}$ and $R_{j,t}$ is
\begin{eqnarray}
{\rm Cov}(R_{i,t},R_{j,t}) = {\rm Cov}(r_{i,t},r_{j,t}) = b_i b_j {\rm
  Var}(M_t).
\label{cov1}
\end{eqnarray}

The covariance between magnitudes of returns depends on the return
distribution of $M_t$ and $R_{i,t}$, but the covariance between squared
magnitudes $r_{i,t}^2$ indicates the properties of the magnitude
cross-correlations. The covariance between $r_{i,t}^2 $ and $r_{j,t}^2 $
is
\begin{eqnarray}
{\rm Cov}(r_{i,t}^2,r_{j,t}^2)=b_i^2 b_j^2 {\rm Var}(M_t^2). 
\label{cov2}
\end{eqnarray}

The above results in Eqs.~(\ref{cov1})-(\ref{cov2}) show that the
variance of the global factor and square of the global factor account
for all the zero time lag covariance between returns and squared
magnitudes.  For time lag covariance between $r_{i,t}$, we find
\begin{eqnarray}
{\rm Cov}(r_{i,t},r_{j,t},\Delta t)&=& E(r_{i,t},r_{j,t-\Delta t})
-E(r_{i,t})E(r_{j,t-\Delta t})\\
&=& b_i b_j A_M(\Delta  t).
\end{eqnarray}
Here 
\begin{equation}
A_M(\Delta t)\equiv E(M_t M_{t-\Delta t})-E(M_t) E(M_{t-\Delta
  t}) 
\end{equation}
is the autocovariance of $M_t$. Similarly, we find
\begin{eqnarray}
{\rm Cov}(r_{i,t}^2,r_{j,t}^2,\Delta t)=b_i^2 b_j^2 A_{M^2}(\Delta t).
\end{eqnarray}
Here 
\begin{equation}
A_{M^2}(\Delta t)=E(M_t^2 M_{t-\Delta t}^2)-E(M_t^2)E(M_{t-\Delta
  t}^2) 
\end{equation}
is the autocovariance of $M_t^2$.

In GFM, the time lag covariance between each pair of
indices is proportional to the autocovariance of the global factor. For
example, if there is short-range autocovariance for $M_t$ and long-range
autocovariance for $M_t^2$, then for individual indices the
cross-covariance between returns will be short-range and the
cross-covariance between magnitudes will be long-range.  Therefore, the
properties of time-lag cross-correlation in multiple time series can be
modeled with a single time series--- the global factor $M_t$.

The relationship between time lag covariance among two index returns and
autocovariance of the global factor also holds for the relationship
between time lag cross-correlations among two index returns and
auto-correlation function of the global factor, because it only need to
normalize the original time series to mean zero and standard deviation one.

\section{Estimation and Analysis of the Global Factor}
\label{ESTIMATION}

\subsection{Estimation of the Global factor}
 
In contrast to domestic markets, where for a given country we can choose
the stock index as an estimator of the ``global'' factor, when we study
world markets the global factor is unobservable. At the world level when
we study cross-correlations among world markets, we estimate the global
factor using principal component analysis (PCA) \cite{Jolliffe86}.

In this section we use bold font for N dimensional vectors or
$N \times N$ matrix, and underscore $t$ for time series.
Suppose ${\bf R_t}\equiv(R_{1,t},R_{2,t},\ldots,R_{N,t})^T$ is the multiple time
series, each row of which is an individual time series
$R_{i,t}=(R_{i,1},R_{i,2},\ldots,R_{i,T})$. We standardize each time
series to zero mean and standard deviation 1 as
\begin{equation}
 z_{i,t}\equiv \frac{R_{i,t}-\langle R_{i,t}\rangle}{\sigma(R_{i,t})}. 
 \label{z}
\end{equation}

The correlation matrix can be calculated as ${\bf C}\equiv \frac{1}{T}
{\bf z_t}{\bf z_t}^T$ where ${\bf z_t}^T$ is the transpose of ${\bf z_t}$, and
the $T$ in the denominator is the length of each time series.  Then we
diagonalize the $N\times N$ correlation matrix ${\bf C}$
\begin{equation}
{\bf C}={\bf U} {\bf \Lambda} {\bf U}^T. 
\end{equation}
Here ${\bf \Lambda}\equiv{\rm diag}(\lambda_1,\lambda_2,..., \lambda_N)$
and $\lambda_1 \geq \lambda_2 \geq ... \geq \lambda_N$ are the
eigenvalues in non-increasing order, ${\bf U}$ is an orthonormal matrix,
whose $i$-th column is the basis eigenvector ${\bf u_i}$ of ${\bf C}$,
and ${\bf U}^{\bf T}$ is the transpose of ${\bf U}$, which is equal to
${\bf U}^{-1}$ because of orthonormality.

For each eigenvalue and the corresponding eigenvector, it holds 
\begin{eqnarray}
\lambda_i={\bf u_i}^T {\bf C} {\bf u_i}={\bf u_i}^T {\rm Cov}({\bf z_t})
 {\bf u_i}={\rm Var}({\bf u_i}^T {\bf z_t})=
{\rm Var}( {\alpha}_{i,t}).  
\label{PCA_value}       
\end{eqnarray}
According to PCA, ${\alpha_{i,t}}={\bf u_i}^T {\bf z_t}$ is
defined as the $i$-th principal component ($\alpha_{i,t}$),
and the eigenvalue $\lambda_i = {\rm Var}({z_{i,t}}) $ indicates the
portion of total variance of ${\bf z_t}$ contributed to
${\alpha_{i,t}}$, as shown in Eq.~(\ref{PCA_value}).  Since the
total variance of ${\bf z_t}$ is
\begin{eqnarray}
\sum_{i=1}^{N}{\rm Var}({z_{i,t}})= {\rm trace}({\bf C})=\sum_{i=1}^{N}\lambda_i,
\end{eqnarray}
the expression $\lambda_i/{\rm trace}({\bf C})$ indicates the percentage
of the total variance of ${\bf z_t}$ that can be explained by the
${\alpha_{i,t}}$. According to PCA (a) the principal components
${\alpha_{i,t}}$ are uncorrelated with each other and (b)
${\alpha_{i,t}}$ maximizes the variance of the linear
combination of ${\bf U}^T {\bf z_t}$ with the orthonormal restriction
${\bf U}^T {\bf U}=1$ given the previous principal components
\cite{Jolliffe86}.

 From the orthonormal property of ${\bf U}$ we obtain
\begin{eqnarray}
 {\bf I}&=&{\bf U U}^T
 ={\bf u}_1 {\bf u}_1^T+ {\bf u}_2 {\bf u}_2^T +...+{\bf u}_N {\bf
   u}_N^T, 
\end{eqnarray}
where {\bf I} is the identity matrix.  Then the multiple time series
${\bf z_t}$ can be represented as a linear combination of all the
$\boldsymbol{\alpha_t}$
\begin{eqnarray}
\nonumber {\bf z_t}&=&({\bf u}_1 {\bf u}_1^T+ {\bf u}_2 {\bf u}_2^T +...+
          {\bf u}_N {\bf u}_N^T){\bf z_t}\\
 &=&{\bf u}_1 {\alpha_{1,t}}+{\bf u}_2
          \alpha_{2,t}+...+{\bf u}_N
          \alpha_{N,t}. 
\label{R2PC}
\end{eqnarray}
The total variance of all time series can be proved to be equal to the
total variance of all principal components
\begin{eqnarray}
\sum_{i=1}^N{\rm Var}({z}_{i,t})&=&{\rm Var}({\bf u}_1) {\alpha_{1,t}}+ 
... + {\rm Var}({\bf u}_N) \alpha_{N,t}\\
&=&\sum_{i=1}^N {\bf u}_i^T {\bf u}_i {\rm Var}(\alpha_{i,t}) = 
\sum_{i=1}^N {\rm Var}(\alpha_{i,t}).
\end{eqnarray}

Next we assume that ${\rm Var}({\alpha_{1,t}})=\lambda_1$ is
much larger than each of the rest of eigenvalues---which means that the
first $\boldsymbol{\alpha_t}$, $ {\alpha}_{1,t}$, accounts for
most of the total variances of all the time series. We express ${\bf z_t}$
as the sum of the first part of Eq.~(\ref{R2PC}) corresponding to
${\alpha_{1,t}}$ and the error term combined from all other
terms in Eq.~(\ref{R2PC}). Thus
\begin{eqnarray}
       {\bf z_t}&=&{\bf u}_1 {\alpha_{1,t}}+{\boldsymbol {\eta_t}}, \nonumber  \\
{\boldsymbol {\eta_t}}&\equiv &\sum_{i=2}^N {\bf u}_i \alpha_{i,t}.
\label{R2PC1}
\end{eqnarray}
Then ${\alpha_{1,t}}$ is a good approximation of the global
factor $M_t$, because it is a linear combination of $R_{i,t}$ that
accounts for the largest amount of the
variance. ${\alpha_1}$ is a zero-mean process because it is
a linear combination of ${z}_{i,t}$ which are also zero-mean
processes (see Eq.~(\ref{z})).

Comparing Eqs.~(\ref{z}) and (\ref{R2PC1}) with 
\begin{eqnarray}
R_{i,t}=\mu_i+b_i M_t+\epsilon_{i,t},
\end{eqnarray}
%
we find the following estimates: 
\begin{eqnarray}
M_t&=&{\alpha_{1,t}}, \nonumber  \\ 
b_i&=&\sigma(R_i)u_{1i}, \nonumber    \\ 
\epsilon_{i,t}&=&\sigma(R_i) \eta_{i,t}. 
 \label{Duan}
\end{eqnarray}
Using Eq.~(\ref{PCA_value}) we find that
\begin{eqnarray}
{\rm Corr}(M_t, R_{i,t})=\sqrt{\lambda_i}u_{i1}.
\label{PC12Cor}
\end{eqnarray}

In the rest of this work, we apply the method of Eq.~(\ref{Duan}) to
empirical data.

\subsection{Analysis of the global factor}

Next we apply the method of Eq.~(\ref{Duan}) to estimate the global
factor of 48 world index returns.  We calculate the auto-correlations of
$M_t$ and $|M_t|$, which are shown in Figs. 3 and 4. Precisely, for the
world indices, Fig.~3(a) shows the time series of the global factor
$M_t$, and Fig.~3(b) shows the auto-correlations in $M_t$.  We find only
short-range auto-correlations because, after an interval $\Delta t=2$,
most auto-correlations in $M_t$ fall in the range of $(-1.96 \sqrt{1/T},
1.96 \sqrt{1/T})$ \cite{Shumway}, which is the 95\% confidence interval
for zero auto-correlations, Here $T=2744$.

For the 48 world index returns, Fig.~4(a) shows the time series of
magnitudes $|M_t|$, with few clusters related to market shocks during
which the market becomes fluctuates more.  Fig.~4(b) shows that, in
contrast to $M_t$, the magnitudes $|M_t|$ exhibit long-range
auto-correlations since the values $|M_t|$ are significant even after
$\Delta t=100$.  The auto-correlation properties of the global factor
are the same as the auto-correlation properties of the individual
indices, i.e., there are short-range auto-correlations in $M_t$ and
long-range power-law auto-correlations in $|M_t|$
\cite{Granger96,Liu}. These results are also in agreement with Fig.~1(b)
where the largest singular value $\lambda_L$ vs. $\Delta t$ calculated
for $\tilde {\textsf{C}}$ decays more slowly than the largest singular
value calculated for \textsf{C}. As found in Ref.~\cite{Podobnik10} 
 for $\Delta t >> 1$,
$\lambda_L(\Delta t)$ approximately follows the same decay pattern as cross-correlation
functions.
Although a Ljung-Box test shows that the return auto-correlation is
significant for a 95\% confidence level \cite{LBtest}, the return
auto-correlation is only 0.132 for $\Delta t=1$ and becomes
insignificant after $\Delta t=2$ .  Therefore, for simplicity, we only
consider magnitude cross-correlations in modeling the
global factor.

We model the long-range market-factor returns {\bf M} with a particular
version of the {\small GARCH} process, the GJR {\small GARCH} process
  \cite{Glosten}, because this {\small GARCH} version explains well the
  asymmetry in volatilities found in many world indices
  \cite{Glosten,Rossiter,Joel10}.  
The GJR {\small GARCH} model can be written as
\begin{eqnarray}
\epsilon_t &=& \sigma_t \eta_t, 
\label{pr2a}\\
\sigma^2_t &=&\alpha_0 + \sum_{i=1}^{q} (\alpha_i + \gamma T_{t-i} )
\epsilon_{t-i}^2 +
\sum_{i=1}^{p} \beta_i \sigma^2_{t-i},
\label{pr2b}
\end{eqnarray} 
where $\sigma_t$ is the volatility and $\eta_t$ is a random process with
a Gaussian distribution with standard deviation 1 and mean 0. The
coefficients $\alpha$ and $\beta$ are determined by a maximum likelihood
estimation
(MLE) and $T_t =1$ if $\epsilon_{t-1} < 0$, $T_t = 0$ if $\epsilon_{t-1}
\ge 0$. We expect the parameter $\gamma$ to be positive, implying that
``bad news'' (negative increments) increases volatility more than ``good
news''.  For the sake of simplicity, we follow the usual procedure of
setting $p=q=1$ in all numerical simulations.  In this case, the
GJR-GARCH(1,1) model for the market factor can be written as
\begin{eqnarray}
M_t &=& \sigma_t \eta_t, 
\label{pr2c}\\
\sigma^2_t &=&\alpha_0 + (\alpha_1 + \gamma T_{t-1} )
\epsilon_{t-1}^2 +
 \beta_1 \sigma^2_{t-1}.
\label{pr2d}
\end{eqnarray} 
We estimate the coefficients in the above equations using MLE, where the
estimated coefficients are shown in Table.~1.

Next we test the hypothesis that a significant percentage of the world
cross-correlations can be explained by the global factor. By using PCA
we find that the global factor can account for 30.75\% of the total
variance.  Note that, according to RMT, only the eigenvalues larger than
the largest eigenvalue of a Wishart matrix calculated by Eq.~(\ref{rmt})
(and the corresponding $\boldsymbol{\alpha}$s) are significant.  To
calculate the percentage of variance the significant
$\boldsymbol{\alpha}$s account for, we employ the {\small RMT} approach
proposed in Ref.~\cite{prl}.  The largest eigenvalue for a
Wishart matrix is $\lambda_+=1.282$ for $N=48$ and $T=2744$ we found in
the empirical data.  From all the 48 eigenvalues, only the first three
are significant: $\lambda_1=14.762$, $\lambda_2=3.453$, and
$\lambda_3=1.380$. This result implies that among the significant
factors, the global factor accounts for $\lambda_1/\sum_{i=1}^{3}
\lambda_i=75.34 \%$ of the variance, confirming our hypothesis that the
global factor accounts for most variance of all individual index
returns.

PCA is defined to estimate the percentage of variance the global factor
can account for zero time lag correlations. Next we study the time lag
cross-correlations after removing the global trend, and apply the
{\small SVD} to the correlation matrix of regression residuals $\eta_i$
of each index [see Eq.~(\ref{duan1})]. Our results show that for both
returns and magnitudes, the remaining cross-correlations are very small
for all time lags compared to cross-correlations obtained for the
original time series.  This result additionally confirms that a large
fraction of the world cross-correlations for both returns and magnitudes
can be explained by the global factor.

\section{Applications of Global Factor Model}
\label{application}

\subsection{Locating and Forecasting global risks}

The asymptotic (unconditional) variance    for the GJR-{\small GARCH} model is
$\alpha_0/(1-\alpha_1-\beta_1-\gamma/2)=10.190$ \cite{Ling}. For the
market factor
the conditional volatility $\sigma_t$ can be estimated
by recursion using the historical conditional volatilities 
 and fitted coefficients in Eq.~(\ref{pr2d}). For example,
the largest cluster at the end of the graph shows the 2008 financial
crisis.  In
Fig.~5(a) we show the time series of the conditional volatility of
Eq.~(\ref{pr2d}) of the global factor.  The clusters in the conditional
volatilities may serve to predict market crashes.  In each cluster, the
height is a measure of the size of the market crash, and the width
indicates its duration. In Fig.~5(b) we show the forecasting of the
conditional volatility of the global factor, which asymptotically
converges to the unconditional volatility.

\subsection{Finding uncorrelated individual indices}
 
Next, in Fig.~\ref{CorMarInd} we show the cross-correlations between the
global factor and each individual index using Eq.~(\ref{PC12Cor}).
There are indices for which cross-correlations with the global factor
are very small compared to the other indices; 10 of 48 indices have
cross-correlations coefficients with the global factor smaller than 0.1.
These indices correspond to Iceland, Malta, Nigeria, Kenya, Israel,
Oman, Qatar, Pakistan, Sri Lanka, and Mongolia. The financial market of
each of these countries is weakly bond with financial markets of other
countries.  This is useful information for investment managers because
one can reduce the risk by investing in these countries during world
market crashes which, seems, do not severely influence these countries.

\section{Discussion}
\label{concl}

We have developed a modified time lag random matrix theory (TLRMT) in
order to quantify the time-lag cross-correlations among multiple time
series. Applying the modified TLRMT to the daily data for 48 world-wide
financial indices, we find short-range cross-correlations between the
returns, and long-range cross-correlations between their magnitudes.
The magnitude cross-correlations show a power law decay with time lag,
and the scaling exponent is 0.25.  The result we obtain, that at the
world level the cross-correlations between the magnitudes are
long-range, is potentially significant because it implies that strong
market crashes introduced at one place have an extended duration
elsewhere---which is ``bad news'' for international investment managers
who imagine that diversification across countries reduces risk.

We model long-range world-index cross-correlations by introducing a
global factor model in which the time lag cross-correlations between
returns (magnitudes) can be explained by the auto-correlations of the
returns (magnitudes) of the global factor. We estimate the global factor
as the first component by
using principal component analysis. Using random matrix theory, we find
that only three principal components are significant in explaining the
cross-correlations. The global factor accounts for 30.75\% of the total
variance of all index returns, and 75.34\% of the variance of the three
significant principle components. Therefore, in most cases, a single
global factor is sufficient.

We also show the applications of the GFM, including locating and
forecasting world risk, and finding individual indices that are weakly
correlated to the world economy. Locating and forecasting world risk can
be realized by fitting the global factor using a GJR-GARCH(1,1) model,
which explains both the volatility correlations and the asymmetry in the
volatility response to both ``good news'' and ``bad news.'' The
conditional volatilities calculated after fitting the GJR-GARCH(1,1)
model indicates the global risk, and the risk can be forecasted by
recursion using the historical conditional volatilities and the fitted
coefficients.  To find the indices that are weakly correlated to the
world economy, we calculate the correlation between the global factor
and each individual index. We find 10 indices which have a correlation
smaller than 0.1, while most indices are strongly correlated to the
global factor with the correlations larger than 0.3.  To reduce risk,
investment managers can increase the proportion of investment in these
countries during world market crashes, which do not severely influence
these countries.

Based on principal component analysis, we propose a general method which
helps extract the most significant components in explaining long-range
cross-correlations. This makes the method suitable for broad range of
phenomena where time series are measured, ranging from seismology and
physiology to atmospheric geophysics.  We expect that the
cross-correlations in EEG signals are dominated by the small number of
most significant components controlling the cross-correlations. We
speculate that cross-correlations in earthquake data are also controlled
by some major components. Thus the method may have significant
predictive and diagnostic power that could prove useful in a wide range
of scientific fields.

We thank Ivo Grosse and T. Preis for valuable discussions and NSF for
financial support.

\eject

\begin{table}
\caption{GJR-GARCH(1,1) coefficients of the global factor.
The P-values and t-values comfirms that all these parameters
are significant at 95\% confidence level.
The positive value of $\gamma$ means ``bad news" has larger impact
on the global market than ``good news".
We find $\alpha_1+\beta_1+\gamma/2=0.9756$, which is very close to
1, and so indicate long-range volatility  auto-correlations.}
\bigskip

\bigskip\bigskip\bigskip        

\begin{tabular}{ccccc}
\hline
\hline
           &    Value&  Std.Error&   t-value&   P-value \\
 $\alpha_0$ & 0.2486 &  0.0283&   8.789& 0.0000\\
 $\alpha_1$ & 0.0170&   0.0080&   2.128& 0.0334\\
  $\beta_1$ & 0.8790&   0.0101&  86.939& 0.0000\\
   $\gamma$ & 0.1591&   0.0148&  10.805& 0.0000\\
\hline
\hline
\end{tabular}
\label{coef}
\end{table}

\begin{figure}[b]
 \includegraphics[width=0.5\textwidth]{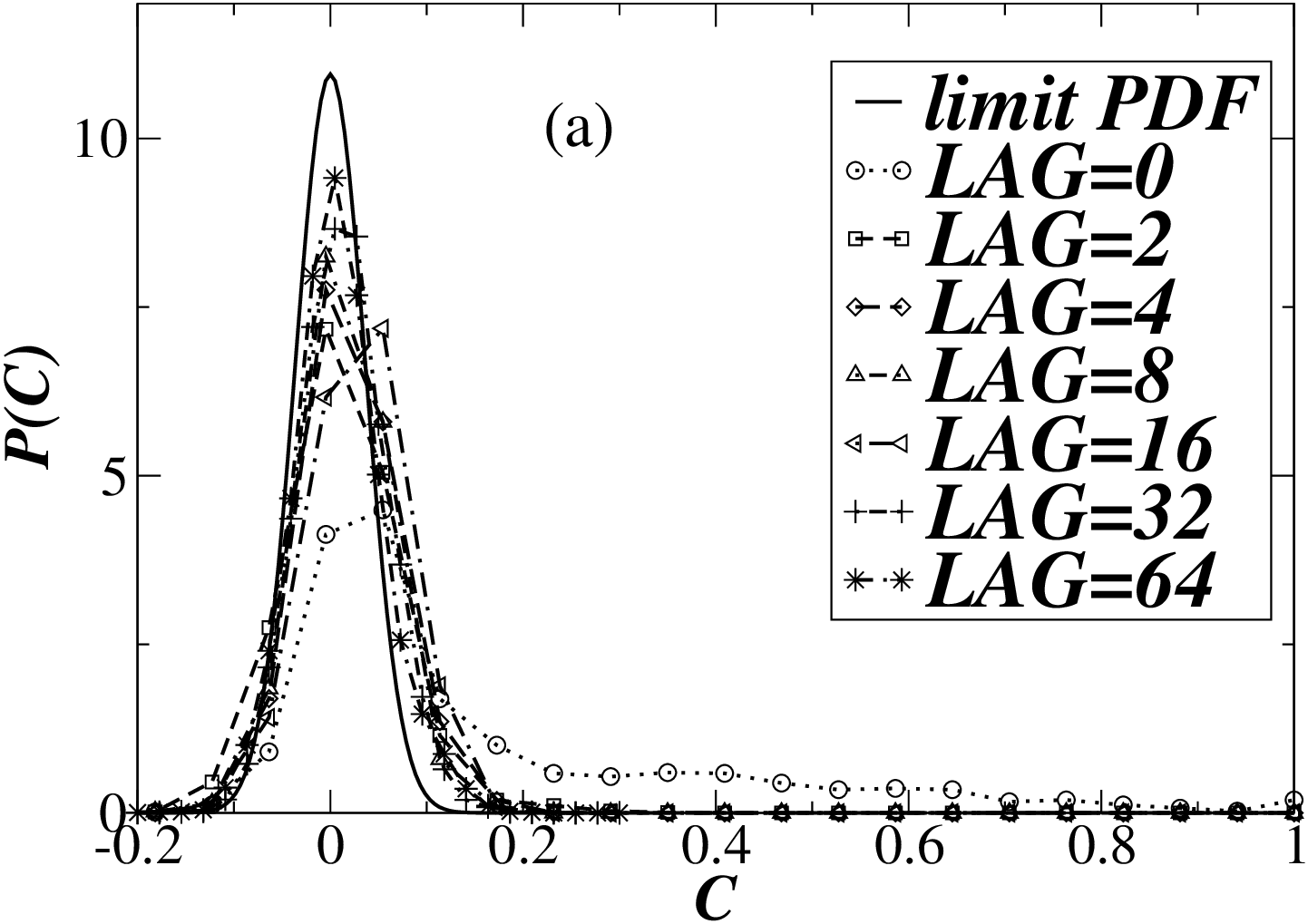}\\
 \includegraphics[width=0.5\textwidth]{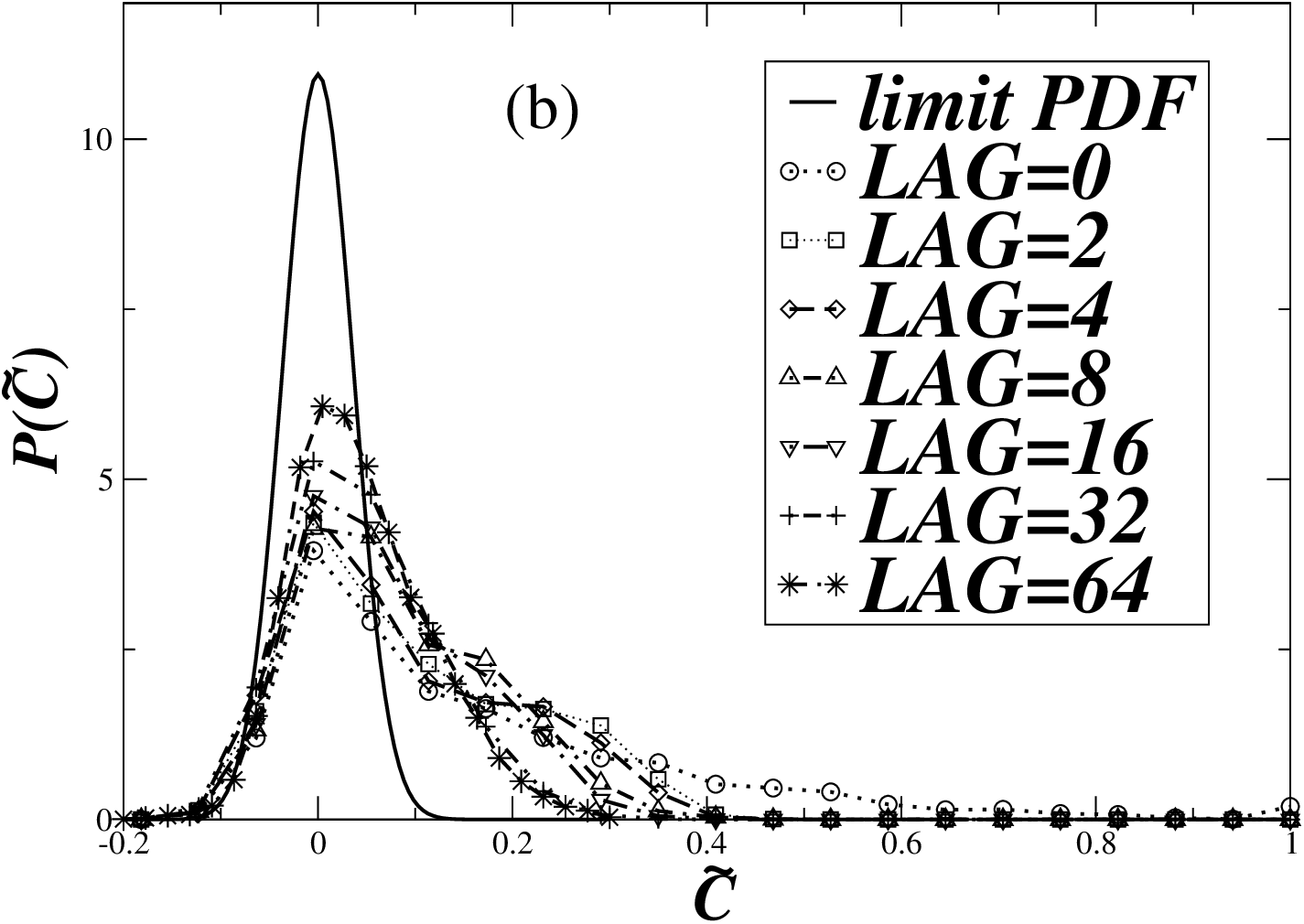} \\
\caption{Cross-correlations among the $N=48$ world financial index
  returns each of size $T=2744$ (a) The empirical pdf of the coefficients
  of the cross-correlation matrix {\textsf{C}} calculated between index
  returns with increasing $\Delta t$ quickly converges to the Gaussian
  form. The normal distribution is the distribution of the pairwise cross-correlations for finite length uncorrelated time series, which 
is a normal distribution with mean zero and standard deviation $\frac{1}{\sqrt{T}}$.
between (b) The empirical pdf of the coefficients of the matrix $\tilde
  {\textsf{C}} $ calculated between index volatilities approaches the
  pdf of the random matrix more slowly than in (a).  }
\label{data1}
\end{figure}

\begin{figure}[b]
 \includegraphics[width=0.6\textwidth]{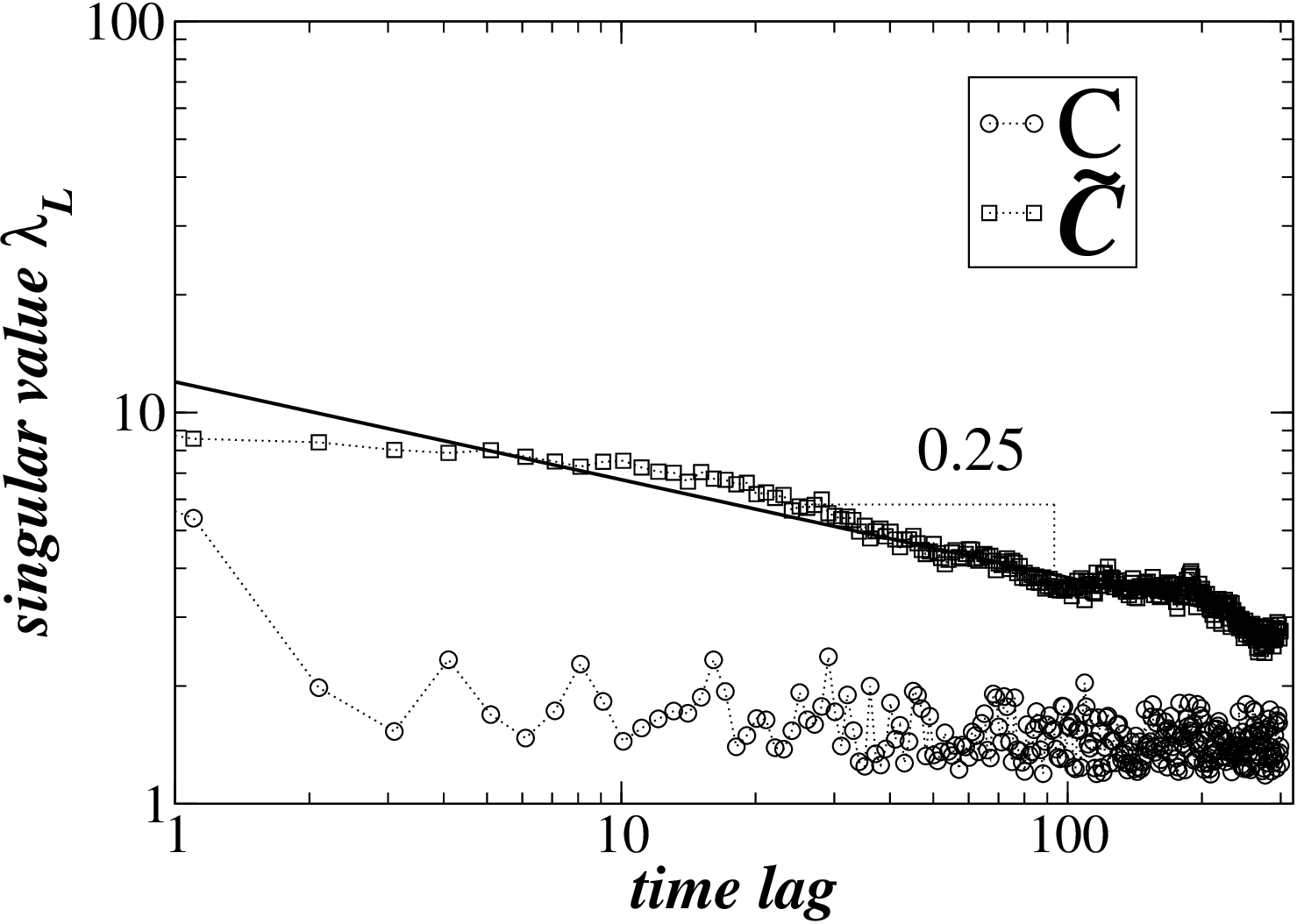}\\
\caption{Long-range magnitude cross-correlations.  The largest singular
  value $\lambda_L$ obtained from the spectrum of the matrices
  \textsf{C} and $\tilde {\textsf{C}} $ versus time lag $\Delta t$.
  With increasing $\Delta t$, the largest singular values obtained for
  \textsf{C} of returns decays more quickly than $\tilde {\textsf{C}} $
  calculated for absolute values of returns.  The magnitude
  cross-correlations decay as a power law function with the scaling
  exponent of $\approx 0.25$.}
\label{SCD}
\end{figure}

 \begin{figure}[b]

\centering \includegraphics[width=0.6\textwidth]{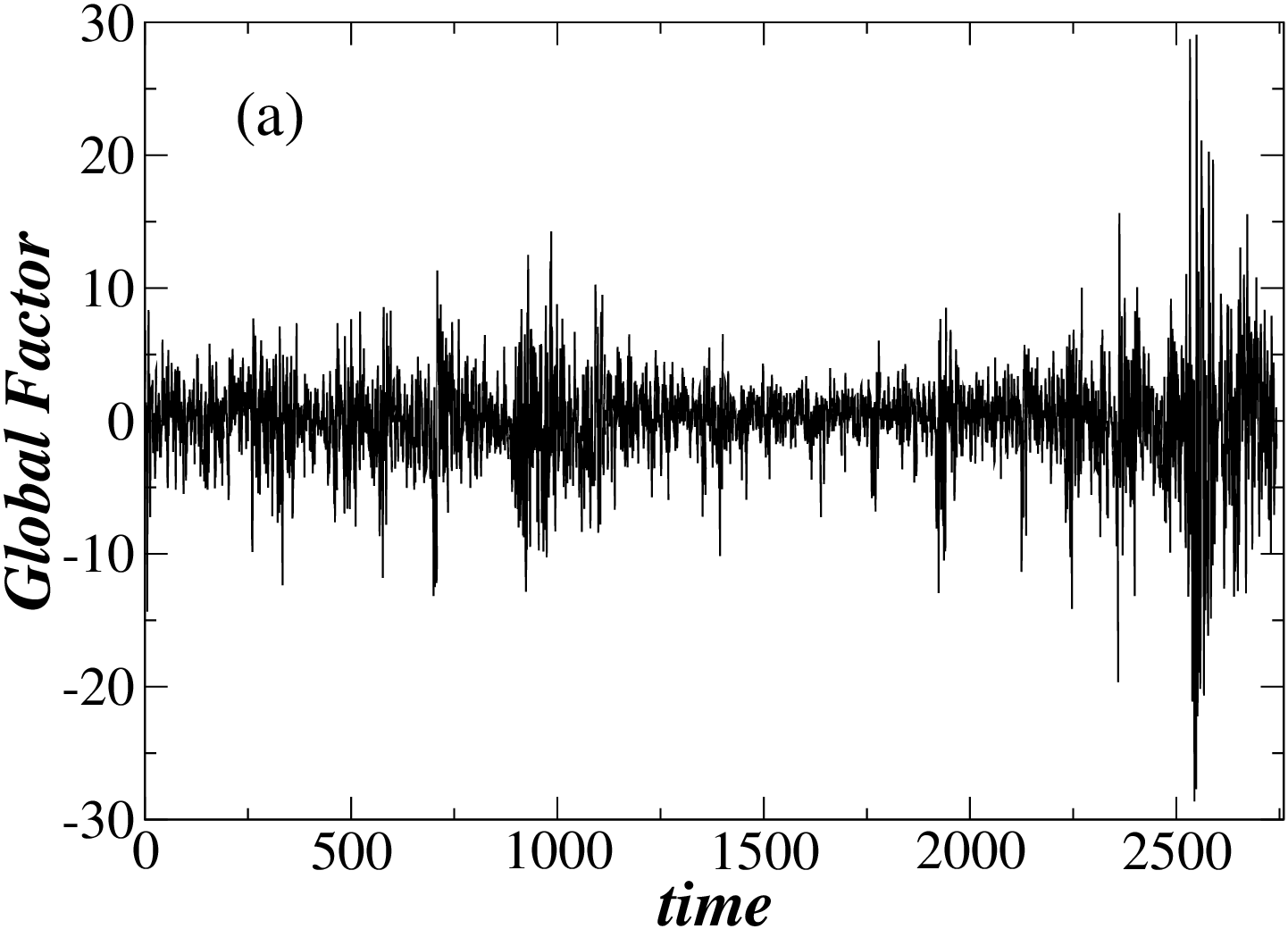}\\
\centering \includegraphics[width=0.6\textwidth]{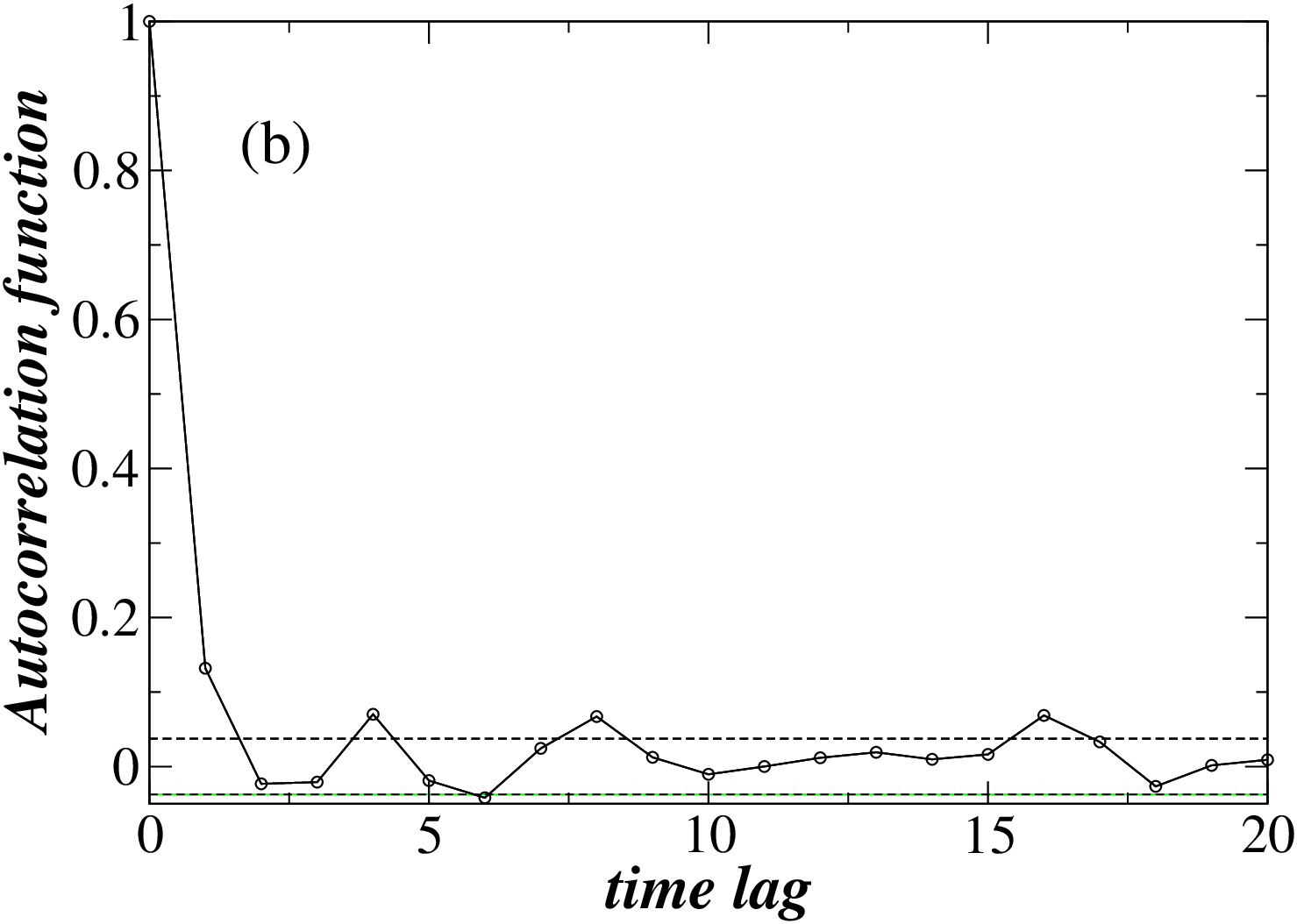}
\vspace{-3.6mm}
\caption{Short-range cross-correlations of a global factor.  (a) Time
  series of the global factor.  (b) The auto-correlation function (ACF)
  of the global factor. The region between dashed lines is the 95\%
  confidence interval for the no auto-correlation hypothesis.
  Auto-correlations are smaller than 0.132 except $\Delta t=0$, and
  become insignificant after time lag $\Delta t=2$, with no more than
  one significant auto-correlation for every 20 time lags. Therefore,
  only short-range auto-correlations can be found in the global factor.}
\label{market1}
\end{figure}

\begin{figure}[b]
\centering \includegraphics[width=0.6\textwidth]{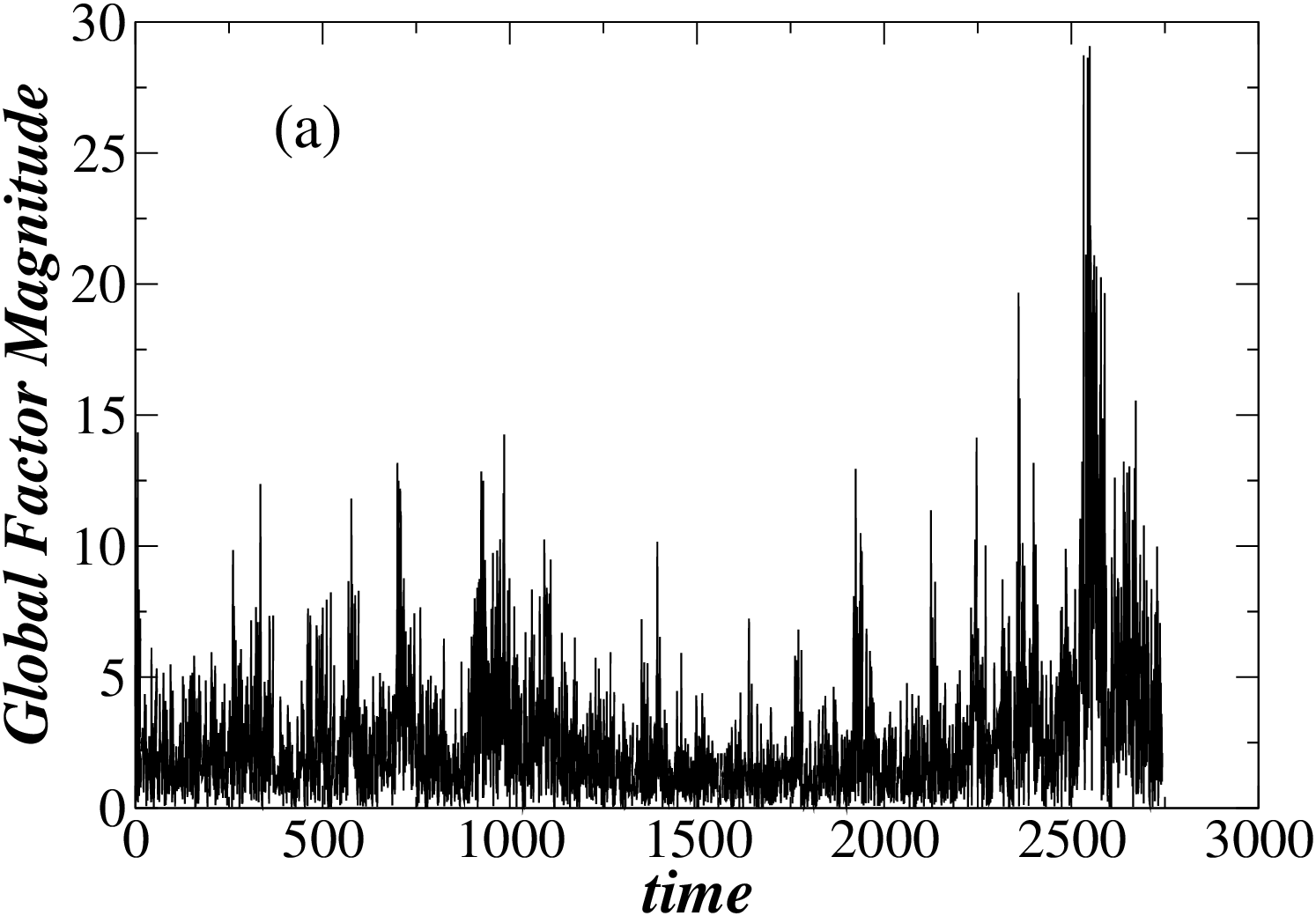}\\
\centering \includegraphics[width=0.6\textwidth]{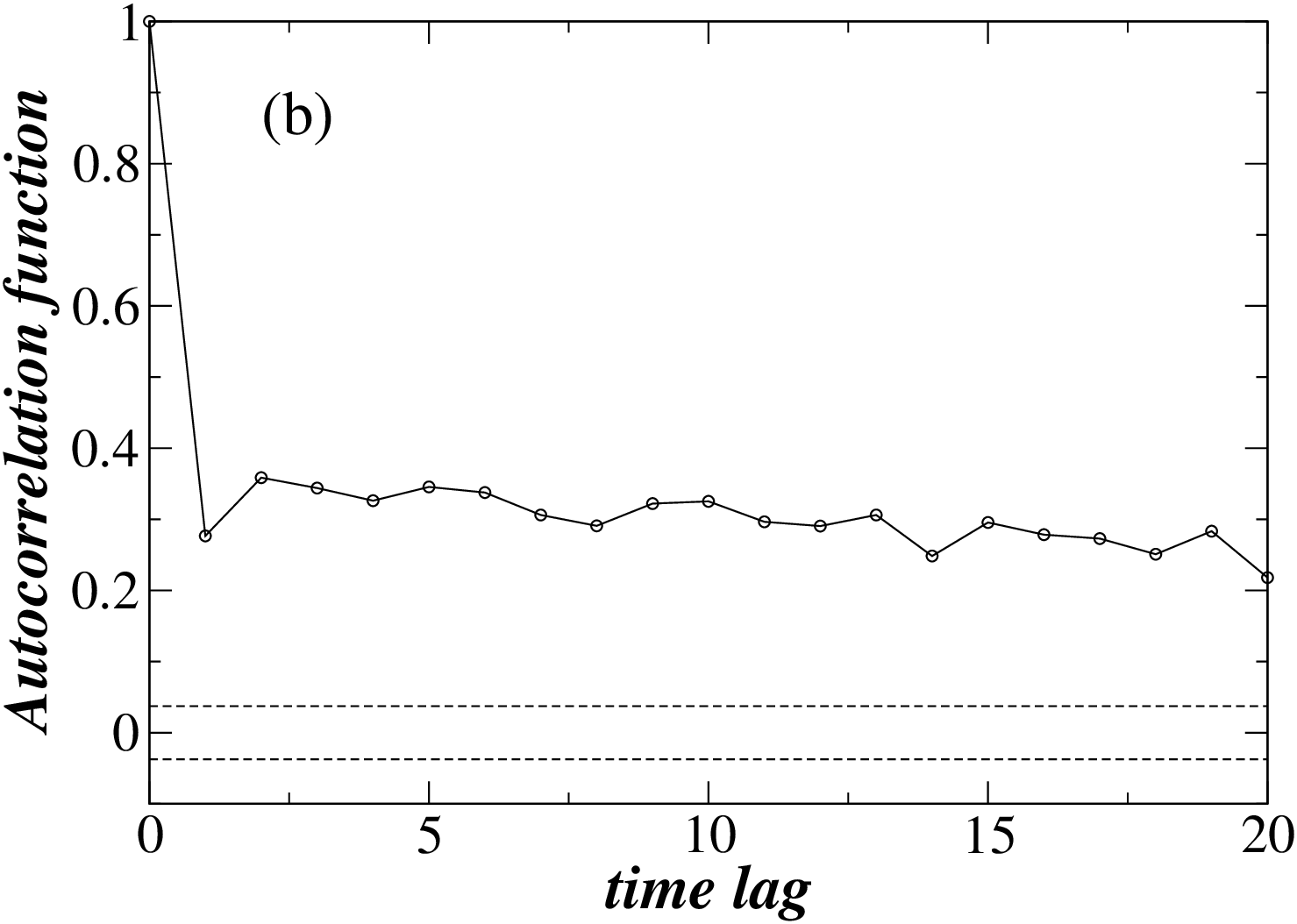}
\vspace{-3.6mm}
\caption{Long-range cross-correlations of the magnitude global
  factor.(a) Time series of magnitudes of the global factor.  (b)
  Auto-correlations of magnitudes of the global factor.  The region
  between dashed lines is the 95\% confidence interval for the no
  auto-correlation hypothesis.  Auto-correlations are much larger than
  the auto-correlations of the global factor itself, is as large as
  0.359 at $\Delta t=2$, and is still larger than 0.2 until $\Delta
  t=33$. For every time lag, the autocorrelation is significant even
  after $\Delta t=100$. Therefore long-range auto-correlations exist in
  the magnitudes of the global factor.}
\label{market2}
\end{figure}

\begin{figure}[b]
\centering \includegraphics[width=0.6\textwidth]{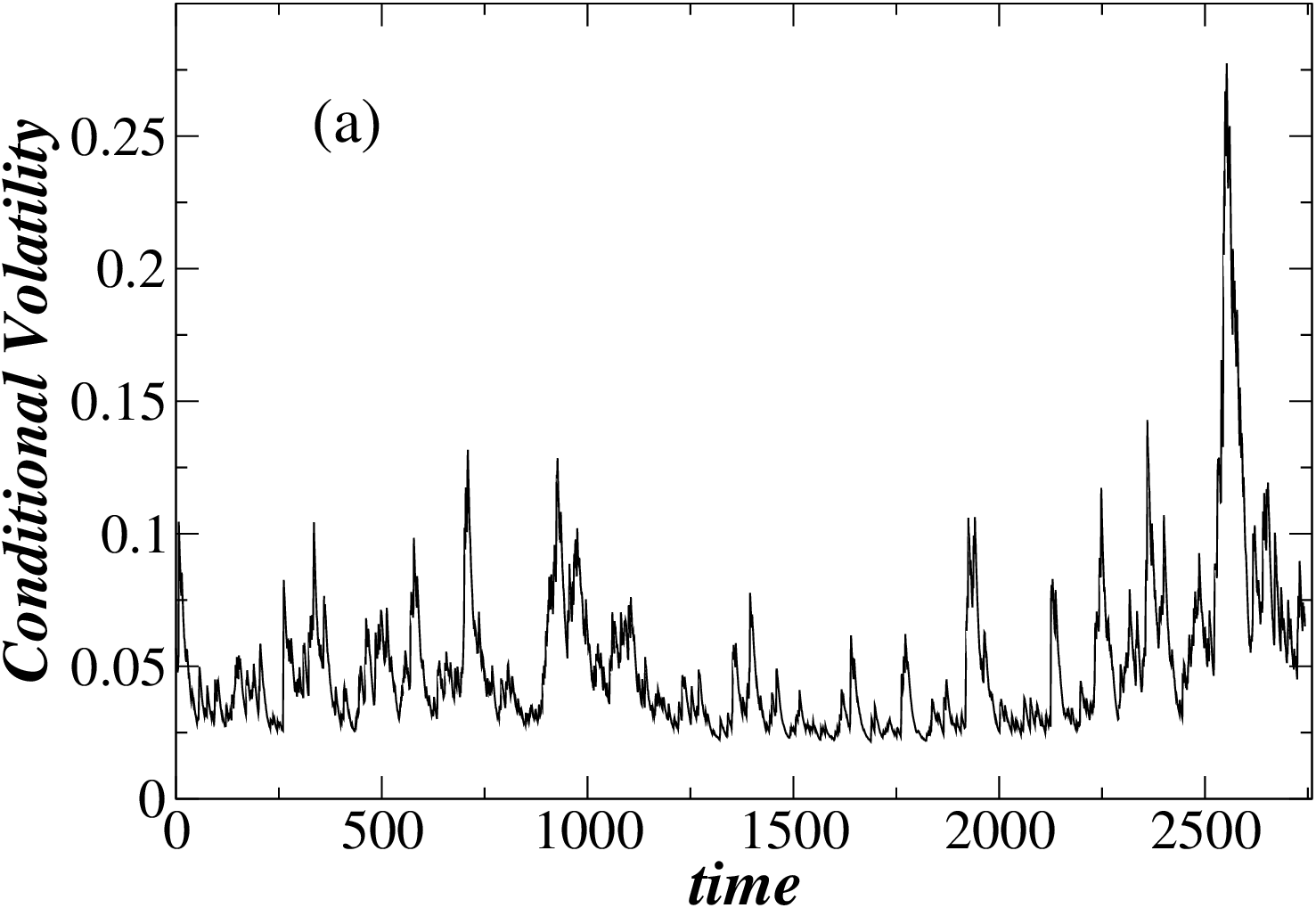}\\
\centering \includegraphics[width=0.6\textwidth]{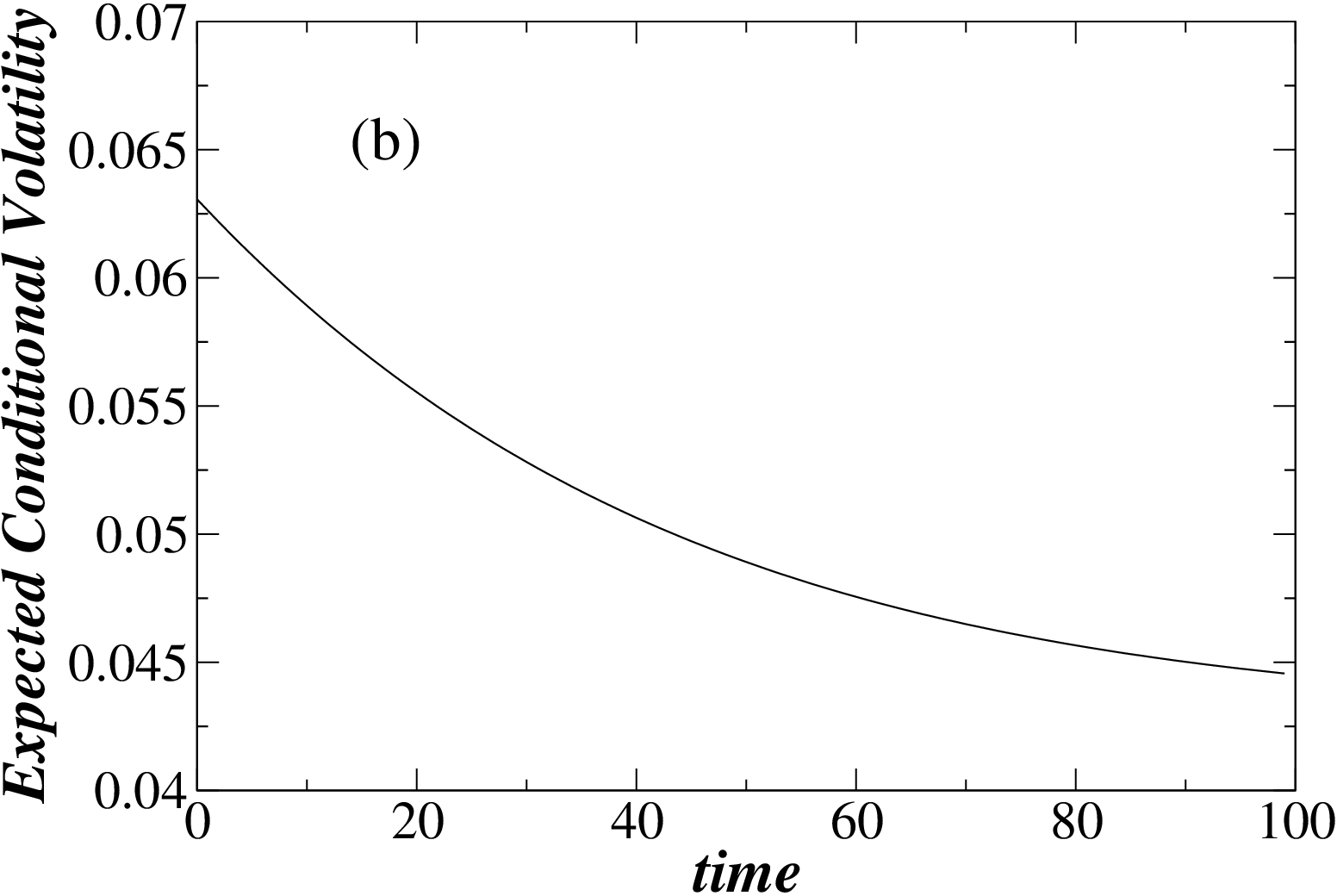}
\vspace{-3.6mm}
\caption{(a) Conditional volatility of the global factor, showing that
  the clusters in the conditional volatilities may serve to predict
  market crashes.  In each cluster, the height indicates the size of the
  market crash, and the width indicates its duration. (b) The 100-day
  forecasted volatility of the global factor, using the past data
  ranging from 4 Jan 1999 through 10 July 2009.  It will converge to the
  unconditional volatility asymptotically.}
\label{cond}
\end{figure}

\begin{figure}[b]
\centering \includegraphics[width=0.6\textwidth]{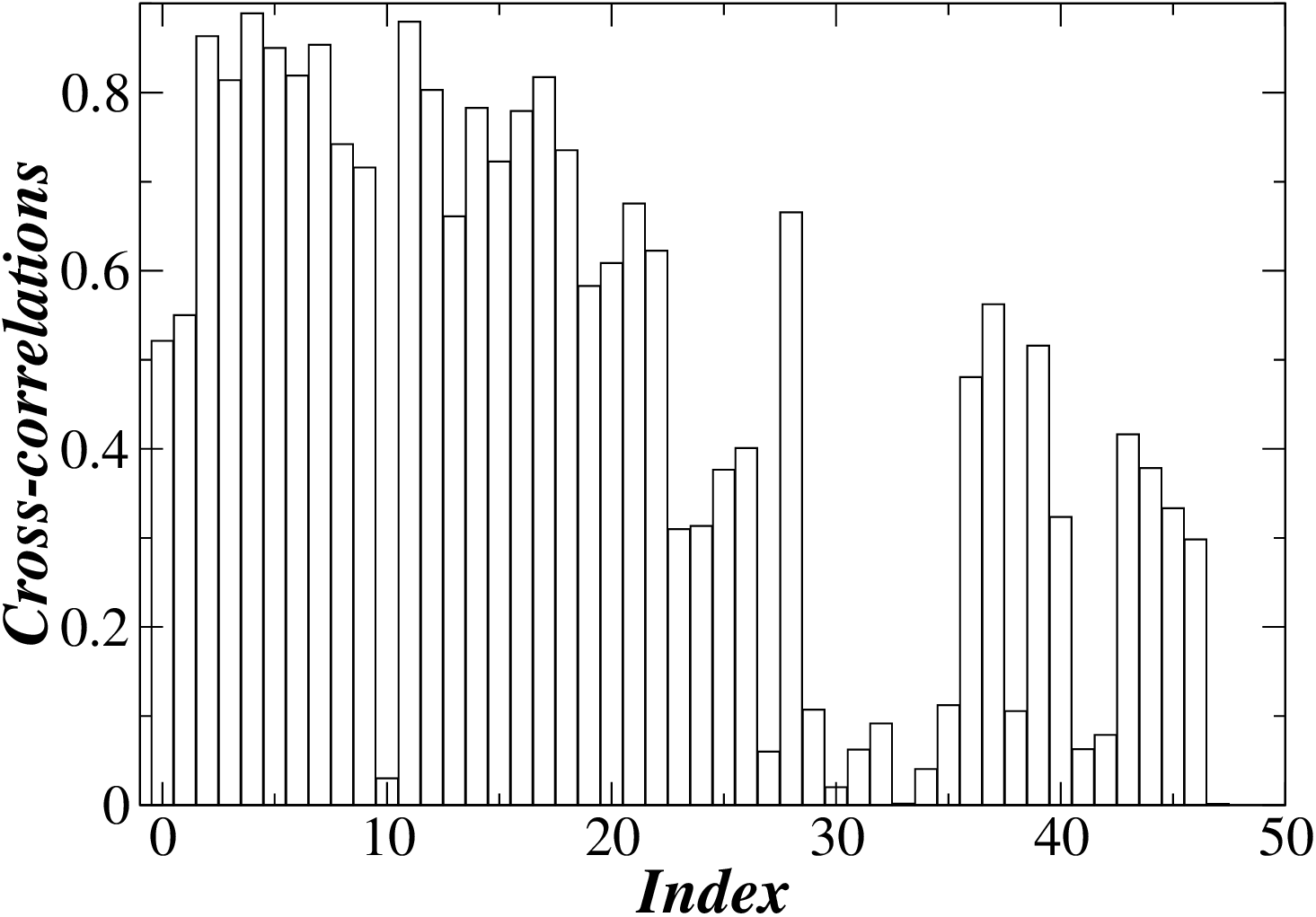}
\caption{Cross-correlation between the global factor $M_t$ and each
  individual index $R_{i,t}$, $i=1,2,...,48$. The global factor has
  large correlation with most of the indices. However, there are indices
  that are not much correlated with the global factor. 10 of the 48
  indices have a correlation smaller than 0.1 between the global factor,
  corresponding to the indices for Iceland, Malta, Nigeria, Kenya,
  Israel, Oman, Qatar, Pakistan, Sri Lanka, and Mongolia.  Hence, unlike
  most countries, the economies of these 10 countries are more
  independent of the world economy.}
\label{CorMarInd}
\end{figure}

\begin{thebibliography}{99}

\bibitem{havlin1}  {R. A. Meyers, ed.,}
{\it Encyclopedia of Complexity and Systems Science}   (Springer,  2009).

\bibitem{chris}  {K. Christensen  and   N. R.  Moloney,}
{\it Complexity and Criticality}   (Imperial College  Press,  2010).

\bibitem{havlin2}  {R. Cohen  and S.  Havlin,}
{\it Complex Networks: Structure, Robustness and Function}
(Cambridge University  Press,  2010).

\bibitem{fn} Heartbeat interval time series are  among many
  time series characterizing the functioning human.

\bibitem{Samuelsson} {P. Samuelsson,   E. V.   Sukhorukov,  and M. Buttiker,}
  {Phys. Rev. Lett.} {\bf 91}, 157002 (2003).

\bibitem{Cottet} { A. Cottet, W.   Belzig,   and  C. Bruder,}
{Phys. Rev. Lett.}  {\bf 92}, 206801 (2004).

\bibitem{Nader} {I. Neder, M. Heiblum,  D. Mahalu,  and  V. Umansky,}
{Phys. Rev. Lett.} {\bf 98}, 036803 (2007).

 \bibitem{Yama} {K. Yamasaki, A.  Gozolchiani,  and S.  Havlin,}
 {Phys. Rev. Lett.} {\bf 100}, 228501 (2008).

  \bibitem{Corral}
A. Corral, Phys. Rev. Lett. {\bf 95}, 159801 (2005).

 \bibitem{Seismology} {M. Campillo  and A. Paul,}
 {Science} {\bf 299}, 547 (2003).

 \bibitem{Lipp}
 E. Lippiello, L. de Arcangelis, and C. Godano, Phys. Rev. Lett.
{\bf 100}, 038501 (2008).

\bibitem{Solnik96} B. Solnik, C. Bourcrelle, and Y. Le Fur,
{Fin. Anal. Journal\/} {\bf 52}, 17--34 (1996).

\bibitem{Erb} C. B. Erb, C. R. Harvey, and T. E. Viscanta,
{Fin. Analysts Journal\/} {\bf 50}, 32--45 (1994).


\bibitem{LeBaron}
B. LeBaron, W.B. Arthur, R. Palmer, J. Econ. Dyn.
Control {\bf 23}, 1487 (1999)


\bibitem{prl} L. Laloux et al., {Phys. Rev. Lett. \/} {\bf 83}, 
  1467-1470 (1999);
V. Plerou {\it et al.}, {\it ibid.} {\bf  83}, 1471--1474 (1999);
{Phys. Rev. E\/} {\bf 65}, 066126 (2002).


\bibitem{takayasubook}  {M. Takayasu  and  H. Takayasu,}  {\it Statistical
  Physics and Economics} (Cambridge University Press, Cambridge, 2010).

\bibitem{Mant99} {  R. N. Mantegna, }  {Eur. Phys. J. B}
 {\bf 11},  193 (1999).


\bibitem{Kertesz} {L. Kullmann, J. Kertesz  and K. Kaski,}  
{Phys. Rev. E} {\bf 66},  026125 (2002). 

\bibitem{Mantegna06} {M. Tumminello, 
 T. Aste,  T.  Di Matteo   and   R. N. Mantegna,}  
{Proc. Natl. Acad. Sci. USA}  {\bf 102},    10421 (2005).


\bibitem{Takayasu06} {T. Mizuno,  H.  Takayasu  and M. Takayasu,}
 {Physica A} {\bf 364},    336 (2006).  


 
 \bibitem{DCCA} B. Podobnik and H. E. Stanley, {Phys. Rev. Lett. \/} 
{\bf 100},   084102 (2008); Podobnik B.  {\it et al.},  
{Proc. Natl. Acad. Sci. USA}  {\bf 106},    22079 (2009).

\bibitem{Carbone}  S. Arianos  and A. Carbone, {J. Stat. Mech.\/} 
{\bf P03037}   (2009).  



 

\bibitem{BP07}
B.  Podobnik  {\it et al.}, 
 European Phys. Journal B  {\bf 56}, 47 (2007).   

 



\bibitem{Mark} H. M. Markowitz,
 {J. Finance\/} {\bf 7}, 77 (1952).


 
 \bibitem{bob10} {P. Bob,  M. Susta, K. Glaslova  and   N. N. Boutros, } 
 {Psychiatry Research} {\bf 177},  37 (2010). 
 
\bibitem{Podobnik10} B. Podobnik {\sl et al.}, 
{Europhys. Lett.}  {\bf 90}, 68001 (2010).

\bibitem{Jolliffe86} I. T. Jolliffe, {\sl Principal Component
  Analysis\/} (Springer-Verlag, New York, 1986).

\bibitem{Guhr03} T. Guhr and B. Kalber,
 {J. Phys. A\/} {\bf 36}, 3009 (2003).

\bibitem{dfa} C. K. Peng  {\it et al}
Phys. Rev. E  {\bf 49}, 1685  (1994);
 A. Carbone, G. Castelli, and H. E. Stanley, 
Physica A {\bf 344}, 267  (2004); 
L. Xu {\it et al},  Phys. Rev. E {\bf 71}, 051101 (2005);
A. Carbone, G. Castelli, and H. E. Stanley,  
Phys. Rev. E {\bf 69}, 026105 (2004). 






\bibitem{Box70} G. E. P. Box and G. M. Jenkins, {\sl Time Series
  Analysis, Forecasting and Control\/} (Holden-Day, San Francisco,
  1970).

\bibitem{Engle82} R. F. Engle,
 {Econometrica\/} {\bf 50}, 987 (1982).





\bibitem{Boll86} T. Bollerslev,
{J. Econometrics\/} {\bf 31}, 307 (1986).

\bibitem{Granger96} Z. Ding and C. W. J. Granger,
 {J. Econometrics\/} {\bf 73}, 185 (1996).



 
 \bibitem{Ashk} Y. Ashkenazy {\it et al.}, 
  {Phys. Rev. Lett. \/} {\bf 86}, 1900   (2001).

\bibitem{Kant} 
J. W. Kantelhardt {\it et al.}, 
{Phys. Rev. E  \/} {\bf 65}, 051908 (2002). 

\bibitem{Livina} 
V. N. Livina  {\it et al.},  {Phys. Rev. E  \/} {\bf 67}, 042101 (2003).  

\bibitem{PREBP05} 
 B. Podobnik   {\it et al.},
 {Phys. Rev. E Rapid Communication  \/} {\bf 71}, 025104(R) (2005). 


\bibitem{Sharpe64} W. F. Sharpe,
 {Journal of Finance\/} {\bf 19}, 425 (1964).

\bibitem{Sharpe70} W. F. Sharpe, {\sl Portfolio theory and capital
  markets\/} (McGraw-Hill, New York, 1970).

\bibitem{Boll88} T. Bollerslev, R. F. Engle, and J. M. A. Wooldridge,
 {The Review of Economics and Statistics\/} {\bf 72}, 498
 (1988).

\bibitem{Boll90} T. Bollerslev,
  {J. Political Economy\/} {\bf 96}, 116 (1990).

\bibitem{Engel95} R. Engle and F. K. Kroner,
 {Econometric Theory\/} {\bf 11}, 122 (1995).


\bibitem{Canarella} G. Canarella and S. K. Pollard,
  {Int. Rev. of Economics\/} {\bf 54}, 445 (2007).

\bibitem{Ding93} Z. Ding, C. W. J. Granger, and R. F. Engle,
{J. Empirical Finance\/} {\bf 1}, 83  (1993).









\bibitem{europa} FTSE 100, DAX, CAC 40, IBEX 35, Swiss Market, FTSE MIB,
  PSI 20, Irish overall, OMX Iceland 15, AEX, BEL 20, Luxembourg LuxX,
  OMX Copenhagen 20, OMX Helsinki, OBX Stock, OMX Stockholm 30,
  Austrian Traded ATX, Athex Composite Share Price, WSE WIG, Prague
  Stock Exch, Budapest Stock Exch INDX, Bucharest BET Index, SBI20
  Slovenian Total Mt, OMX Tallin OMXT, Malta Stock Exchange IND,
  FTSE/JSE Africa TOP40 IX

\bibitem{asia} ISE National 100, Tel Aviv 25, msm30, dsm 20, Mauritius
  Stock Exchange, NIKKEI 225, Hang Seng, Shanghai se b share, all
  Ordinaries, nzx all, Karachi 100, Sri Lanka Colombo all sh, Stock Exch
  Of Thai, Jakarta Composite, FTSE Bursa Malaysia KLCI, PSEi -
  Philippine SE, MSE Top 20

\bibitem{amer} S\&P 500, Mexico BOLSA

\bibitem{africa} FTSE/JSE Africa TOP40 IX, CFG 25, Nigerian Stock
  Exchange All Shares, Nairobi Stock Exchange 20-Share

\bibitem{Longin95} F. Longin and B. Solnik,
 {J. International Money and Finance\/} {\bf 14}, 3
 (1995).
 

\bibitem{Ramchand} L. Ramchand and R. Susmel,
 {J. Empirical Finance\/} {\bf 5}, 397      (1998).
\bibitem{Mehta91} M. L. Mehta, {\sl Random Matrices\/} (Academic Press,
  Boston, 1991).

\bibitem{guhr} T. Guhr, A. M\"uller-Groeling, and H. Weidenm\"uller, 
 {Physics Reports\/} {\bf 299}, 190 (1998).

\bibitem{laloux99b} L. Laloux, P. Cizeau,  J.-P. Bouchaud, and
  M. Potters,
  {Risk\/} {\bf 12}, 69 (1999).

\bibitem{Utsugi} A. Utsugi, K. Ino, and M.  Oshikawa,
{Phys. Rev. E\/} {\bf 70}, 026110 (2004).

\bibitem{pan} R. K. Pan and S. Sinha,
 {Phys. Rev. E\/} {\bf 76}, 046116 (2007).

\bibitem{shen} J. Shen and B. Zheng,
{Europhys. Lett.}  {\bf 86}, 48005 (2009).

\bibitem{Pott05} M. Potters et al., 
 {Acta Phys. Polonica B\/} {\bf 36}, 2767 (2005).

\bibitem{new2} J. Kwapien et al., 
 {Acta Phys. Polonica B\/} {\bf 37}, 3039 (2006).

\bibitem{Bou07} J.-P. Bouchaud et al., 
 {Eur. Phys. J. B} {\bf 55}, 201 (2007).


\bibitem{Gupta} A. M. Sengupta and P. P. Mitra,
 {Phys. Rev. E\/} {\bf 60}, 3389 (1999).

\bibitem{Shumway} R. H. Shumway and D. S. Stoffer, {\sl Time Series
  Analysis and Its Applications\/} (Springer-Verlag, New York, 2000).


\bibitem{Liu} Y. Liu {\sl et al.} 
  {Physica A\/} {\bf 245}, 437 (1997);
   {Phys. Rev. E\/} {\bf 60}, 1390 (1999); P. Cizeau et al., 
Physica A {\bf 245}, 441  (1997).


\bibitem{Ross} S. A. Ross,
{J. Economic Theory\/} {\bf 13}, 341 (1976).

\bibitem{Sharpe63} W. F. Sharpe,
{Management Science\/} {\bf 9}, 277 (1963).


\bibitem{LBtest} G. M. Ljung and G. E. P. Box,
{Biometrika\/}   {\bf 65}, 297  (1978).

\bibitem{Glosten} L. Glosten, R. Jagannathan, and D. Runkle,
 {J. Finance\/} {\bf 48}, 1779 (1993).

\bibitem{Rossiter} R. Rossiter and S. A. Jayasuriya,
 {J. Int. Finance and Eco.}  {\bf 8}, 11 (2008).


\bibitem{Joel10} J. Tenenbaum {\sl et al.}, 
  {Phys. Rev. E\/} {\bf 82}, 046104 (2010).











\bibitem{Ling} S. Ling and M. McAleer,
 {J. Econometrics\/} {\bf 106}, 109 (2002).















\end{thebibliography}
\end{document}